\documentclass{article}
\usepackage{fullpage,amsmath,amsthm,amssymb,graphicx,charter,color,subfigure}
\usepackage{url}
\title{Exploration and retrieval of \\whole-metagenome sequencing samples}
\author{
Sohan Seth\,$^{1}$, 
Niko V\"{a}lim\"{a}ki\,$^{2,3}$,
Samuel Kaski\,$^{1,3}$, 
Antti Honkela\,$^{3}$\\
$^{1}$Helsinki Institute for Information Technology HIIT,\\
Department of Information and Computer Science,\\ 
Aalto University, Espoo, Finland\\
$^{2}$Genome-Scale Biology Program and Department of Medical Genetics,\\
University of Helsinki, Helsinki, Finland\\
$^{3}$ Helsinki Institute for Information Technology HIIT,\\
Department of Computer Science, University of Helsinki, Helsinki, Finland
}
\begin{document}

\maketitle
\renewcommand{\SS}[1]{#1}
\newcommand{\AH}[1]{#1}
\begin{abstract}

Over the recent years, the field of whole metagenome shotgun sequencing has
witnessed significant growth due to the high-throughput sequencing technologies
that allow sequencing genomic samples cheaper, faster, and with better coverage
than before. This technical advancement has initiated the trend of sequencing
multiple samples in different conditions or environments to explore the
similarities and dissimilarities of the microbial communities. Examples include
the human microbiome project and various studies of the human intestinal
tract. With the availability of ever larger databases of such measurements,
finding samples similar to a given query sample is becoming a central
operation.  
In this paper, we develop a content-based exploration and retrieval method for
whole metagenome sequencing samples.  We apply a distributed string mining
framework to efficiently extract all informative sequence $k$-mers from a pool
of metagenomic samples and use them to measure the dissimilarity between two
samples.  \SS{We evaluate the performance of the proposed approach
on two human gut metagenome data sets as well as human microbiome
project metagenomic samples.  We observe significant enrichment
for diseased gut samples in 
results of queries with another diseased sample and
very high accuracy in discriminating 
between different body sites even though the method is unsupervised.}
A software implementation of the DSM framework is available at
\url{https://github.com/HIITMetagenomics/dsm-framework}.

\end{abstract}

\section{Introduction}

Metagenomics is the study of microbial communities in their natural habitat
using genomics techniques~\cite{tyson_community_2004}. It is undergoing a boom
due to proliferation of high-throughput sequencing technologies.  Many studies
focus at targeted sequencing of specific marker genes such as the 16S rRNA gene
in bacteria, but recently there has been a growing interest in whole metagenome
sequencing (see, e.g.~\cite{qin_human_2010,consortium_structure_2012}).  While
targeted studies provide data for phylogenetic profiling at a lower cost, whole
metagenomes provide much more information, for example, about the collective
metabolism~\cite{Greenblum2012}, and the population genetics of the
community~\cite{Schloissnig2013}.  Recent studies have also found associations
between features of whole human gut metagenomes and type II
diabetes~\cite{Qin2012}.  New data are accumulating rapidly, with a popular
web-based MG-RAST server~\cite{meyer_metagenomics_2008} listing almost 3000
public whole metagenomes.

Analysing whole-metagenome shotgun (WMS) sequencing data is very challenging.
The original sample typically contains genetic material from hundreds to
thousands of bacterial species of different abundances~\cite{Li2012}, most of
which have not been fully sequenced previously.  After sequencing, we obtain a
huge collection of short sequence reads whose species of origin is unknown.
While significant progress has been made, analysis relying on either the
limited previously annotated genomes, or assembling the reads into novel more
complete genomes, remains difficult and inefficient, and potentially
susceptible to annotation biases.

In this paper we introduce an efficient purely data-driven feature extraction
and selection method as well as similarity measures for WMS sequencing data
sets, and apply them in retrieval of similar data sets.  Such content-based
retrieval is an extremely powerful tool for exploration of the data and
generating hypotheses of disease associations, as previously demonstrated with
gene expression data~\cite{Caldas09,Caldas2012}. Retrieval from existing
databases makes it possible to automatically explore a much greater variety of
hypotheses than relying solely on the more common specifically designed focused
studies.

Content-based similarity measures and retrieval of similar metagenomic data
sets have been suggested previously \cite{Mitra2009,Liu2011,Su2012,jiang_comparison_2012}, based on
quantifying abundances over a relatively small number of predetermined features
requiring existing annotation. Up to some thousands of known taxa, genes or
metabolic pathways have been used. We introduce similarity measures that are
based solely on raw sequencing reads, and hence, unbiased and insensitive to
the quality of the existing annotation. A similar measure has been previously
suggested by \cite{Maillet2012}, but only for pairwise comparisons using a
method that is computationally too expensive to scale to even modestly large
data sets.  Furthermore, instead of considering all
sequences of particular length, also known as $k$-mers, as has been done
earlier for other tasks and by \cite{Maillet2012},
we employ an efficient distributed string mining
algorithm to find informative subsequences that can be of \emph{any} length.

\begin{figure}[t]
\centering
\includegraphics[scale=1]{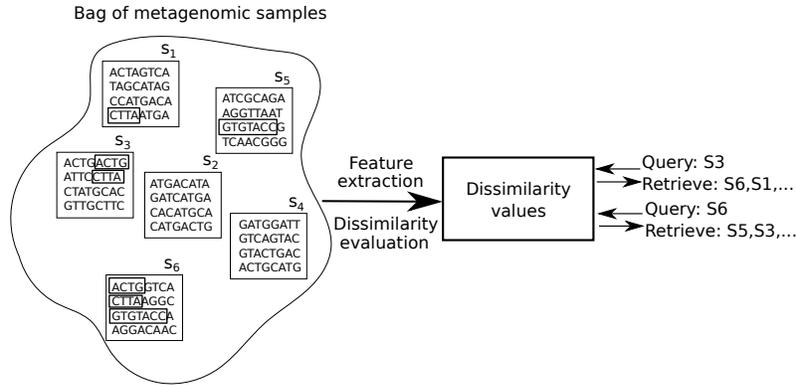}
\caption{Given a set of metagenomic samples our objective is to be
  able to retrieve relevant samples to a query sample. For this, we 
  need to extract relevant features and evaluate a pairwise similarity
  (or dissimilarity) measure. The samples are then ranked in the order
  of increasing dissimilarity from the query.}
\label{fig:metret}
\end{figure}

In order to deal with the very large number of features some feature selection
is necessary.  Previous approaches for detecting relevant features in
metagenomic data have been based on direct comparison of two classes of
samples.  Again, most of these methods work on up to some thousands of
features~\cite{White2009,Parks2010,Segata2011}, with the notable exception of
one study~\cite{Qin2012} where quantification and association testing was done
for over 4.3 million predefined genes.  Without feature selection one can use
short $k$-mers~\cite{baran_joint_2012} or limit to a set of $k$-mers that are
likely to be informative, such as $k$-mers associated with well characterised
protein families~\cite{Edwards2012}.  While there are no previous examples of
unsupervised feature selection for metagenomics, it is a common practice in
information retrieval with text documents~\cite{Yang1997}; a particularly
relevant method assesses the entropy of the distribution of documents in which
a specific term occurs~\cite{Largeron2011}.

We evaluate the performance of the proposed unsupervised, unconstrained
retrieval method on synthetic data, as well as metagenomic samples from human
body sites \cite{qin_human_2010,Qin2012,consortium_structure_2012}.  \SS{To
evaluate the performance of the retrieval engine, we use external
validation based on a ground truth
similarity between two samples. To simplify this process, we consider
a binary similarity, which is crude but easily accessible.} The human gut samples in
\cite{qin_human_2010,Qin2012} come from studies exploring the change in
bacterial species composition between healthy persons and either inflammatory
bowel disease or type II diabetes.  \SS{We utilize disease state to construct a
binary ground truth. Thus, we} study if, given the metagenomic sample of
a person with a disease, the retrieval finds metagenomic samples related by
having the same disease. \SS{In the body site data
\cite{consortium_structure_2012} we use the body sites as
ground truth to investigate whether it is possible to identify
the bacterial communities at different body sites in an unsupervised
setting without the need of reference genomes.
It should be noted that especially for the gut data, two samples may
be related in other ways too. The external validation with
one simple ground truth nonetheless provides an objective platform
for comparing different methods.} Given that the method is
unsupervised and hence completely oblivious of the disease labels, if such
retrieval is successful it is a promising starting point for developing methods
for leveraging data from earlier patients in early detection of disease and
personalized medicine.
 
\begin{figure}[t]
\centering
\includegraphics[scale=1]{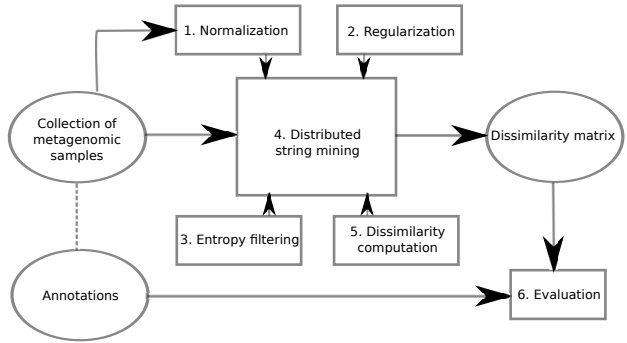}
\caption{Processing steps of our method. Given a collection
  of metagenomic samples, we use the collection as an input to the
  distributed string mining method (4). For the method, we estimate
  the frequency of each $k$-mer (1,2), evaluate if the $k$-mer is
  informative or not (3), and compute the needed dissimilarities
  (5). Finally, in this paper we evaluate the performance considering
  the existing annotations as ground truth; annotations are not needed
  for the retrieval in general.}
\label{fig:pipeline}
\end{figure}

\section{Approach}

Our objective is to extract and select suitable features for representing WMS
sequencing samples and to form a pairwise dissimilarity measure for a
collection of such samples. Given this dissimilarity one can query with a
sample and retrieve other samples that are similar to it
(Fig.~\ref{fig:metret}). The measure needs to be reasonably rapidly computable,
yet captures relevant differences between the samples, and does all this with as
little prior biological knowledge and annotations as possible, since detailed
quantitative prior knowledge is typically not yet available for metagenomics.

Evaluating dissimilarity requires representing the metagenomic sample in a
suitable feature space.  A standard choice for representing objects over
strings is to estimate the $k$-mer frequency values, where a $k$-mer here is a
string of $k$ letters from the DNA alphabet $\{$A,C,T,G$\}$.  Therefore, there
are $4^k$ possible $k$-mers for any given $k$.  It is standard practice to set
$k$ to a specific value, typically a small value to keep the estimation problem
tractable both computationally and statistically. A larger $k$ would give
better discriminability but not without bounds, as for finite data set sizes
there simply is not enough data to estimate long $k$-mers. We argue that
instead of setting $k$ to a particular value, it is more effective to estimate
all possible $k$-mers for all possible $k$ which the data supports.  This makes
the problem more challenging, since the number of such observed different
$k$-mers for large $k$ becomes very large and they become more susceptible to
sequencing errors.  Focusing on $k$-mers appearing more than once in a sample
helps significantly because it is relatively rare to have the exactly same
sequencing errors in two independent reads.

To make the method computationally efficient we treat each $k$-mer as an
independent feature. We compute a Bayesian estimate of their relative
frequencies across samples. The employed prior helps in suppressing noise
caused by small observed read counts.  In the \emph{filtering} step the
abundance distribution of each $k$-mer over samples is used to judge
informativeness of the $k$-mer for retrieval; a $k$-mer with constant abundance
does not have discriminative power and, in the other extreme, a $k$-mer which
is present in only one sample cannot generalize over samples. We show that the
filtering step significantly improves the retrieval performance \SS{with
most datasets and distance measures}.  Finally, we compute the dissimilarity
between two samples across the features as a weighted average of distances
between relative frequencies of individual $k$-mers.  Treating each $k$-mer as
an independent feature allows us to execute these steps fast and on the fly
without storing the intermediate results. Such simplified distance measures are
necessary to guarantee scalability given the extremely high dimensionality of
the $k$-mer features.

To summarize, we introduce methods to i.~estimate the frequencies of a large
number of $k$-mers over multiple samples, ii.~decide if a $k$-mer is
informative or uninformative in the context of a retrieval task, iii.~compute a
distance metric using the filtered $k$-mer frequencies, and iv.~execute these
steps fast without explicitly storing the frequency values.
Fig.~\ref{fig:pipeline} summarizes the method.

\begin{figure}
\centering
\includegraphics[width=\textwidth]{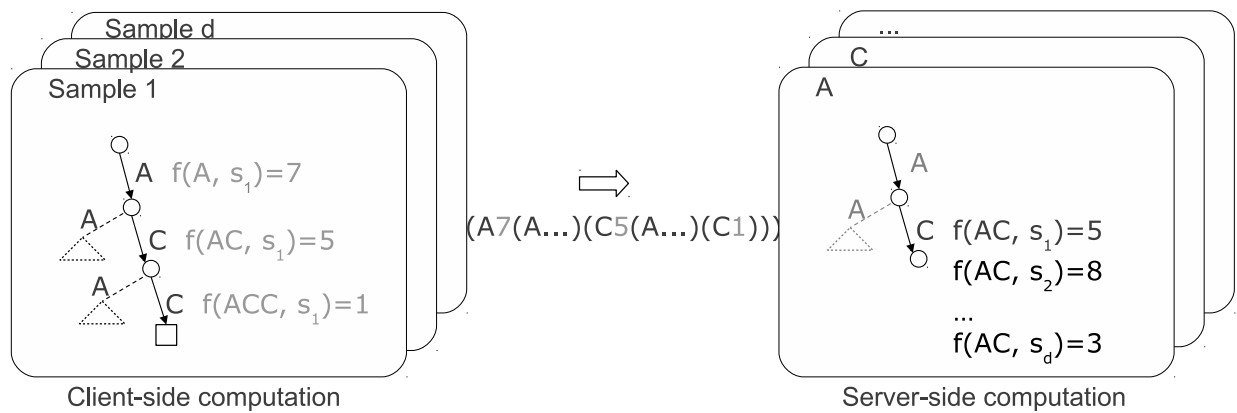}
\caption{
  Technical overview of our
  distributed string mining framework consisting of
  client (left) and server (right) processes.  The client-side
  processes are responsible for computing the substring frequencies
  within each sample $s_1,s_2,\ldots s_d$ separately. Substrings and
  their frequencies are found using a depth-first-traversal over a
  (compressed) suffix tree. Frequency information is transmitted over
  to the server-side by streaming it as a balanced-parenthesis
  representation of a sorted trie. For example, the trie on the left results as
  the parenthesis representation given in the middle.  The server reads the
  client-streams and merges the (already sorted) tries in recursive
  manner: at each node, the server computes the entropy based on the
  received values and updates the affected pairwise distances.
  Load-balancing on the server-side is achieved by hashing the prefix
  of the substring so that each server corresponds to a certain range
  of hash values.}
\label{fig:dsm}
\end{figure}

\section{Methods}

\subsection{Estimating $k$-mer frequencies: normalization, 
  regularization, filtering}

In order to perform the feature selection or filtering, we first compute
Bayesian estimates of the relative frequencies $p(s|w)$ of each $k$-mer $w$
over samples $s \in S$ using observed frequencies $\hat{f}(s, w)$ of the
$k$-mers. These are distributions over samples for each $k$-mer that are
computed independently for each $k$-mer for reasons of computational
efficiency.

Even if the relative abundance of a $k$-mer is the same in every sample, the
observed frequencies may differ because of different sequencing depth or
coverage in different samples.  To tackle this issue we employ normalization:
we normalize the frequency $\hat{f}(s,w)$ by a sample-specific constant
$\sigma(s)$, which is proportional to the total number of base pairs in a
sample, and $\sigma(s)=1$ for the largest sample in the collection in terms of
total base pair count, obtaining
\begin{equation}
  \label{eq:f}
  f(s,w)=\hat{f}(s,w)/\sigma(s).
\end{equation}
The $\sigma(s)$ can be interpreted probabilistically as the probability of
observing a sequence in the actual sample, assuming every sample had the same
number of base pairs to start with but some have been lost in the processing.

In order to estimate the relative frequencies, we place a conjugate symmetric
Dirichlet prior on the parameters of the multinomial distribution over the
observed counts.  The common choice of uniform prior distribution corresponds
to a Dirichlet distribution with all parameters equal to $1$.  This yields a
posterior mean estimate of the relative frequency values as
\begin{align}\label{eq:p_s_w}
p(s|w)=\frac{f(s,w) + 1}{\sum_{s' \in S} \left[f(s',w) + 1\right]}.
\end{align}
The Dirichlet prior with all parameters equal to one is ubiquitous in document
retrieval.  It is particularly suitable for metagenomics due to the following
observations:
\begin{enumerate} 
\item The distributed string mining algorithm (described below) trades off low
$k$-mer counts for speed and ignores any $k$-mers that are present only once
in a sample. The pseudo-count from the prior makes up for this missing count.  
\item Adding pseudo-counts assists in playing down the significance of very
rare $k$-mers that may appear due to sequencing errors in the filtering step
without affecting other $k$-mers too much.
\end{enumerate}

Finally, given the massive number of potential $k$-mers, it is crucially
important to improve signal-to-noise ratio by focusing on the informative ones.
For the unsupervised tasks of comparing the samples, obviously only $k$-mers
which distinguish between the samples are informative. As a concrete example,
consider a $k$-mer that is present in all samples with a similar abundance. It
certainly does not give information useful for comparing samples. In the other
extreme, if a $k$-mer is present in one specific sample, but not in any other,
it is potentially a spurious $k$-mer due to sequencing error, and in any case
does not help in comparing samples either.  On the other hand, if a $k$-mer is
present in some samples, but not all, then it gives information that those
samples are similar in a specific sense.  Informativeness in this sense can be
measured by the entropy $H$ of the distribution of the $k$-mer over the
samples: we filter the $k$-mers based on the conditional entropies
\begin{equation}
  \label{eq:entropy_def}
  H(S|w) = -\frac{1}{\log(|S|)}\sum_{s \in S} p(s|w)\log p(s|w) ;
\end{equation}
a $k$-mer is taken into account in distance computation only if the normalized
entropy is lower than a certain threshold $e$. By
design $0 \leq H \leq 1$. 
\SS{Notice that in standard \emph{information theory}
terminology higher entropy implies higher information.  However, in our context
an informative $k$-mer has low entropy.} Also, due to the Bayesian estimation, a
spurious $k$-mer having only very small counts will have large conditional
entropy and will be filtered out. 

\SS{The optimal value of threshold $e$ varies
with datasets. It can be `optimized' in a supervised manner by utilizing a
\emph{training set} where we have labelled samples. In the absence of a
labelled set, we suggest taking the `average' of distance metrics computed over
the potential thresholds as the final metric. We refer to the final metrics in the two
cases as \textit{optimized metric} and \textit{average metric}.  In our
experimental set-up, we randomly make a 50-50 split of a given dataset in
training $S_{\mathrm{tr}}$ and testing $S_{\mathrm{te}}$ sets:
$S_\mathrm{tr}\cap S_{\mathrm{te}}=\emptyset$, and $S_\mathrm{tr}\cup
S_{\mathrm{te}}=S$. We use $S_{\mathrm{tr}}$ to optimize the entropy threshold:
we query with samples in $S_\mathrm{tr}$, and retrieve relevant samples within
the same set to observe which entropy threshold results in the best 
retrieval result (see Sec.~\ref{sec:evaluation-metric} for details). While
comparing the performance of two methods we always present the evaluation
over $S_{\mathrm{te}}$: we query with samples within
$S_{\mathrm{te}}$, and we retrieve relevant samples from $S$ (not just $S_{\mathrm{te}}$).}

\subsection{Algorithms to extract informative $k$-mers}

Our main computational challenge is to extract all informative $k$-mers from
large-scale datasets in feasible time and space.  Recall that the filtering
step relies on knowledge over multiple samples to decide if the respective
$k$-mer is informative for the retrieval task or not. Since the typical
collections of WMS samples are huge in size, we cannot assume that even the
plain input fits into the main memory of any single machine.  To process these
large-scale datasets, the computation needs to be done either using external
memory (i.e. disk) or in a distributed manner (i.e. a computer cluster).  We
review two approaches: {\em $k$-mer counting}
\cite{marcais_fast_2011,rizk_dsk:_2013} and {\em distributed string mining}
\cite{distmining}. The first one is a standard approach in the literature for
fixed $k$, but has several limitations when applied in our context of multiple
samples and large-scale data. We show that the latter approach is more flexible
in this context and can also be generalized to extract informative $k$-mers
over all values of $k$ simultaneously.

Jellyfish \cite{marcais_fast_2011} and DSK \cite{rizk_dsk:_2013} are examples
of recent algorithmic improvements in $k$-mer counting.  Both tools use hash
tables to compute the $k$-mer distribution for a given (fixed) $k$. In both
tools, space-efficiency is achieved by keeping most of the hash table on disk.
The main drawback with these disk-based approaches is that they are aimed at
counting $k$-mers in a single sample and extending them over to multiple
samples is non-trivial. For example, Jellyfish could, in principle, be extended
to count $k$-mers over multiple samples: the authors give a roughly linear time
algorithm to merge two or more hash tables. However, the intermediate $k$-mer
counts would need to be stored on disk, which requires significant amount of
additional space, and the merge-phase is not parallelized \cite[User manual,
Sect. Bugs]{marcais_fast_2011}.

The decision whether a particular $k$-mer is informative or not is made by
looking at its frequency over all the given WMS samples.  We tackle this
problem by a \emph{Distributed String Mining} (DSM) framework \cite{distmining}
that can handle multi-sample inputs by utilizing a computer cluster.  The main
advantages of this framework are that (i) load-balancing divides the data and
computation over multiple cluster nodes, (ii) intermediate $k$-mer counts are
not stored explicitly, and (iii) there is no additional disk I/O strain, except
reading through the input once. \SS{These advantages allow terabyte-scale data
analysis on a cluster consisting of nodes having limited main memory.}  We
extend the DSM framework to be compatible with our definition of informative
$k$-mers (see the above subsection). It allows us to extract the informative
$k$-mers either for a fixed $k$ or over all values of $k$ in feasible time.

The DSM framework is based on a client-server model. The clients have
one-to-one correspondence to the given samples, each client being responsible
for computing the frequencies within the designated sample.  The client-side
computation relies heavily on {\em suffix sorting} techniques and on
space-efficient data structures for strings \cite{distmining}: the input data
are first preprocessed into a compressed representation, which replaces the
input data and acts as an efficient search structure. The server-side
computation is more straightforward: the server simply merges the (sorted)
input from the clients, computes the entropies and updates the distance
matrices.  Fig.~\ref{fig:dsm} gives a toy example of the client-server
interaction. Two crucial observations are needed to keep the whole computation
and transmission costs feasible. First, the informative $k$-mers can be seen as
a subset of {\em left-right-branching} substrings, i.e. substrings whose
instances have differentiating continuation on both left and
right. \SS{More formally: substring $w$ of string $T[1,n]$ is called {\em
right-branching} if there exists two symbols $a$ and $b$ such that $a\neq b$
and both $wa$ and $wb$ are substrings of $T$. Similarly, a substring $w$ is
{\em left-branching} if $aw$ and $bw$, $a\neq b$, are substrings of $T$.  If a
substring is both left-branching and right-branching we say it is {\em
left-right-branching}.} Second, for any string of length $n$, there are at
most $O(n)$ left-right-branching substrings, and the total length of all such
substrings is bounded by $O(n \log n)$ \cite[Theorem 1]{kmp2009}.

The first observation allows us to reduce the client-side computation to a
smaller set of substrings: it is easy to see that if $k$-mer $w$, having
frequency $f'(s,w)\geq 2$, is non-branching, then there exists a substring $w'$
of length $k'>k$ that is left-right-branching and has exactly the same
frequency, i.e., $f'(s,w)=f'(s,w')$. It follows that the frequency of
non-branching $k$-mers can be deduced from the branching $k'$-mers, and the
left-right-branching substrings contain all the necessary information for us to
detect informative $k$-mers. The second observation guarantees a feasible
transmission cost between clients and servers: the upper bound for the
concatenation of all left-right-branching substrings also acts as an upper
bound for both the server-side running time and the amount of communication
needed.  The drawback of restricting to left-right-branching substrings is that
the informative $k$-mers that we are able to detect have to appear at least
twice in a sample, although this limit may be useful in pruning spurious
$k$-mers introduced by sequencing errors.  More detailed explanation and
analysis of the DSM framework is given in \cite{distmining}.
A software implementation of the DSM framework is available at
\textrm{https://github.com/HIITMetagenomics/dsm-framework}.

\subsection{Dissimilarity metrics}

Having extracted the informative $k$-mers, we use them to compute the
dissimilarity between two metagenomic samples.  We consider three dissimilarity
metrics that can be computed easily over a large number of $k$-mers in
sequential manner, i.e. one k-mer at a time, and without storing all the k-mer
frequencies explicitly. To utilize the natural variance structure of the
$k$-mers|some are more abundant than others|we weight the relative frequencies
of each $k$-mer by their respective total counts, i.e., we utilize the absolute
frequencies $f(s,w)$ as defined in \eqref{eq:f}.

We mainly use the very simple Jaccard distance which does not consider
abundances at all, only whether a $k$-mer occurs or not. Given two sets $s_1$
and $s_2$ of $k$-mers detected as present in two different samples, Jaccard
distance measures how many elements are shared between these two sets.
Mathematically, it is defined as 
\[ 
D_{\mbox{count}}(s_1,s_2) = 1 - \frac{|s_1\cap s_2|}{|s_1 \cup s_2|}.  
\] 
Despite its simplicity, we observe that Jaccard distance performs well; a
potential reason is its robustness to measurement noise and effectiveness when
two metagenomic samples differ in terms of presence and absence of certain
species or functionalities. We assume a $k$-mer is present in a sample if its
frequency is more than 2.

We also experiment with two metrics that use the abundance information:

I. \emph{Variance-stabilized Euclidean distance:} An obvious distance measure
between two metagenomic samples $s_1$ and $s_2$ is the Euclidean distance
between their respective $k$-mer frequencies.  We consider the distance metric
\[
D_{\mbox{sqrt}}(s_1,s_2) = \sum_w(\sqrt{f(w,s_1)} - \sqrt{f(w,s_2)})^2
\]
which can be computed sequentially as new informative $k$-mers are extracted.
The square root transformation is the \emph{variance stabilizing
transformation} for Poisson distribution|a popular model for quantitative
sequencing data.

II. \emph{Log transformed Euclidean distance:} We also consider the same metric
but with log transformation which is a popular approach in document retrieval,
i.e.,
\[
D_{\mbox{log}}(s_1,s_2) = \sum_w (\log(1+f(w,s_1)) - \log(1+f(w,s_2)))^2.
\]
The motivation for using the $\log$ transformation is that it decreases
sensitivity to high frequency counts: some $k$-mers are present in high
abundance in almost every genome, for instance $k$-mers from the marker gene,
and the $\log$ transformation reduces their effect in the metric.

\subsection{Evaluation metric}
\label{sec:evaluation-metric}

We evaluate the performance of the dissimilarity metric in terms of its
performance in the task of retrieving relevant samples given a query
metagenomics sample. \SS{The ground truth for relevance is either the disease class
(disease vs not) or the known body site: 
samples from the same class are considered relevant.}
\begin{table}[t]
\centering
\begin{tabular*}{\textwidth}{@{\extracolsep{\fill}}llrrrr}
 & & HIGH-C & MetaHIT & T2D-P2  & HMP \\
\hline
\multicolumn{2}{l}{Input size (GB)} & 149 & 536 & 786 & 3,353 \\
\multicolumn{2}{l}{Samples}         & 200 & 124 & 199 & 435 \\
\multicolumn{2}{l}{Preproc. (h)}    & 0.4 & 3.6 & 10 & 65 \\
\multicolumn{2}{l}{Total memory (GB)} & 117 & 209 & 610 & 2,885 \\
\hline
All $k$ & Wall-clock (h) & 4.9 & 2.0 & 8.0 & 53 \\
        & CPU time (h) & 149 & 187 & 1,137 & 20,000 \\
\hline
$k=21$ & Wall-clock (h) & 1.8 & 0.4 & 2.8 & 12 \\
       & CPU time (h) & 10 & 74 & 279 & 4,000 \\
\end{tabular*}
\caption{Computational resources required by the distributed string
  mining on different datasets. We report wall-clock times
  and total CPU times for both fixed $k=21$ and over all $k$.
  Preprocessing is done only once, separately from the actual
  computation. Total memory is the memory requirement over all computation nodes. 
  Experiments were ran on a cluster of Dell PowerEdge
  M610 nodes having 32 GB of RAM and 16 cores. 
  Simulated data and MetaHIT were run using up to 8 nodes. T2D-P2 was ran 
  using 32 nodes allowing more parallelization at the server-side. 
  \SS{HMP was ran on a cluster of 20 nodes with 2x10-core Xeon CPUs and 256GB RAM.}}
\label{table:cpu}
\end{table}

For measuring retrieval performance we use an evaluation metric which is
popular in document retrieval, the \emph{mean average precision} (MAP)
\cite{Smucker:2007}. Given a query $q$, the retrieval method ranks the samples
in an increasing order of their dissimilarities from $q$. Given one has
retrieved the top (closest) $n \in \{1,\ldots,N\}$ samples the precision @$n$
is defined as
\begin{align*}
\mathrm{Precision}(n;q) = \frac{\text{number of relevant samples in } n
  \text{ retrieved samples}}{n},
\end{align*}
and MAP defined using \emph{average precision} as,
\begin{equation*}
  \mathrm{MAP}=\frac{1}{|Q|}\sum_{q \in Q}\mathrm{AveP}(q),\; 
\mathrm{AveP}(q)=\frac{1}{m_q}\sum_{n \in R_{q}} \mathrm{Precision}(n;q).
\end{equation*}
Here $Q$ is the set of all queries, $m_q$ is
the number of relevant samples to query $q$, and $R_q$ is the set of locations
in the ranked list where a relevant sample appears. It is straight-forward that
a higher MAP implies better performance.  To judge if two MAP values are
significantly different or not, we employ the randomization test described in
\cite{Smucker:2007}: \SS{for each query, this test randomly reassigns the AvePs
achieved by two methods to one another and computes the difference between the resulting MAP
for multiple such reassignments to get a distribution, against which the true
MAP value is tested in terms of p-value}. In case two samples share the same dissimilarity from
a query sample, we employ the modification suggested in
\cite{mcsherry_computing_2008} to break ties. When computing the mean, we
follow a leave-one-out cross-validation type approach using each sample as a
query, and retrieving from the rest of the collection. \SS{For simulated data and
human gut samples, we only query with the positive samples in the testing set
$q \in S_\mathrm{te}$, whereas for body site samples we query with each sample
in the testing set. For both cases we retrieve from the entire set $S\setminus\{q\}$.
While choosing the entropy threshold in a supervised setting, we query from
$q \in S_{\mathrm{tr}}$ and retrieve from $S_\mathrm{tr}\setminus \{q\}$.}

\subsection{Synthetic data generation}
\label{sec:synth}
\SS{
To test the method, we simulated four datasets containing samples from separate
classes, with the interpretation that samples from the same class are relevant.
In all the datasets we have two classes: both classes of samples have the same 
species composition but different relative abundances. 
We used MetaSim \cite{richter_metasimsequencing_2008} to generate Illumina
reads of length 80 using the error configuration file provided by the
developers. 
Each dataset contains 200 samples: 98 of them belong to the \emph{positive} class 
and the rest belong to the \emph{negative} class.
For each dataset, we used the same 100 species from the following genera:
acetobacter, acetobacterium, acidiphilium, acidithiobacillus, acinetobacter, bacillus,
bacteroides, bifidobacterium, chlamydia, chlamydophila, clostridium, escherichia,
haloarcula, halobacterium, lactobacillus, pasteurella, salmonella, staphylococcus, and
streptococcus.
The abundance profiles were generated from two Dirichlet distributions; one for
positive and the other for negative class.  The parameters of the Dirichlet
distributions were shared between two classes: for half of the species
(randomly chosen) the same parameters were used for both classes and for the
other half of the species the parameters were randomly permuted. For example,
given 5 species the assigned parameters could be: $(0.3,0.2,0.6,0.1,0.9)$ and
$(0.9,0.2,0.3,0.1,0.6)$ where the parameters for the second and fourth species are
the same, but for the other species they were permuted. The exact species and
corresponding parameter values can be downloaded from \url{https://github.com/HIITMetagenomics}.
The resulting datasets are:
\begin{enumerate}
\item HIGH-C: relatively easy data with high coverage (10,000,000 reads per sample)
\item LOW-C: relatively difficult data with low coverage (2,000,000 reads per sample)
\item MIXED-C: mixed data with half the samples from HIGH-C and the rest from LOW-C
to simulate varying sequencing depth.
\item HIGH-VAR: relatively difficult data with same coverage as $\mathrm{HIGH}$ 
but additional noise in the class distributions to simulate more overlap 
between classes. To elaborate, the relative abundance of species is 
$p_{\mathrm{HIGH-VAR}} = 0.5\,p_{\mathrm{HIGH}} + 0.5\,noise$
where $noise$ is generated from a symmetric Dirichlet distribution with 
all parameters equal to 1. 
\end{enumerate}
}

\begin{figure*}[t]
\centering
\includegraphics[width=0.75\textwidth]{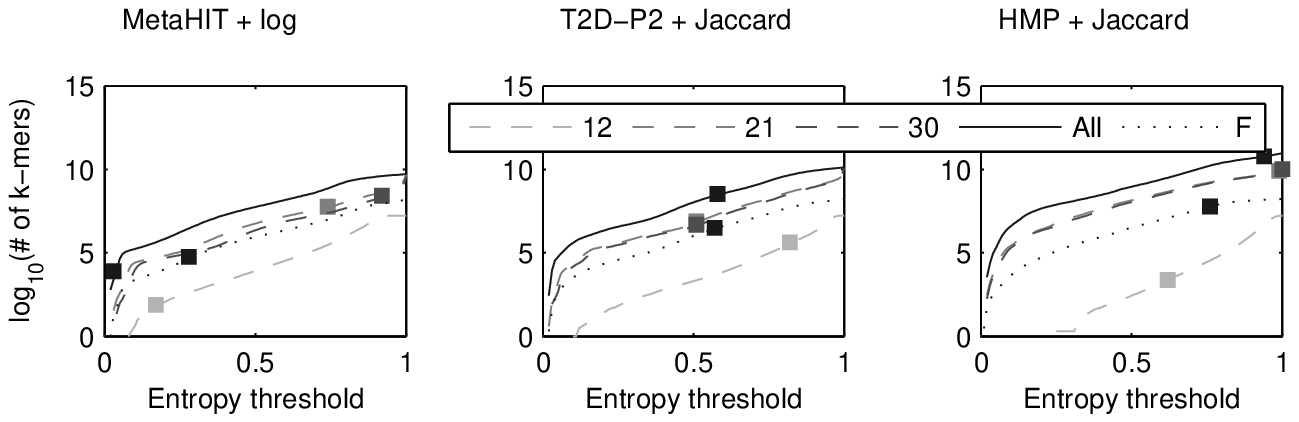}
\includegraphics[width=\textwidth]{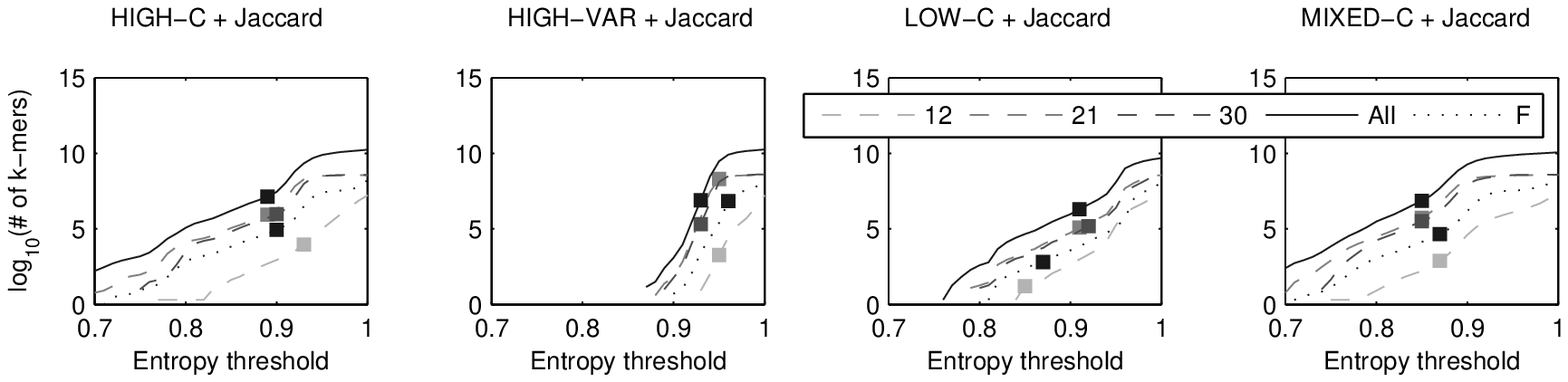}
\caption{\SS{Number of informative strings over varying entropy thresholds
for the proposed approach `All', fixed $k$-mer lenthgs `12',`21' and `30', and
for protein family based comparison with FIGfam `F'. The box denotes the
`optimized' entropy threshold that has been used to evaluate the performance
of the methods.
Some general observations are as
follows. The number of strings for $k=12$ is lower than the rest while the
number of strings for `All' is much higher than rest of the methods, and number
of strings for $k=21$ and $k=30$ are very close. We observe that there 
are strings with low entropies---more in the real data sets than in 
the simulated data sets---which indicate the presence of discriminative
features. Also, the `optimized' entropy threshold varies for different methods.
}}
\label{fig:entth}
\end{figure*}

\section{Results}

\SS{
We evaluated the retrieval performance on three human metagenomics datasets:}
\begin{enumerate} 
\item MetaHIT \cite{qin_human_2010}, 124 metagenomic
samples from 99 healthy people and 25 patients with inflammatory bowel disease
(IBD) syndrome. Each sample has on average $65 \pm 21$ million reads. Our goal
was to retrieve IBD positive patients. 
\item T2D Phase II \cite{Qin2012}, 199 metagenomic samples from 100 healthy
people and 99 patients with type II diabetes. Each sample has on average $47
\pm 11$ million reads. Our goal was to retrieve diabetic patients. We chose to
explore the phase II data instead of the phase I data since the former has
higher coverage; about 40$\%$ more reads than the latter. 
\item \SS{HMP \cite{consortium_structure_2012}, 435 metagenomic samples from 10
different body sites. Out of 690 samples that passed the QC assessment 
(http://www.hmpdacc.org/HMASM/), we discarded 255 samples that had
less than 1\% of the number of reads of the largest sample.}
\end{enumerate}
\SS{To recapitulate, for MetaHIT and T2D-P2, our goal is to observe if given a
positive sample, e.g., from a patient with a particular disease, one can
retrieve relevant samples, i.e., with similar disease; 
whereas for HMP, our goal is to observe
if given a sample from a particular body site, one can retrieve relevant samples,
i.e., samples from the same body site.}
For all data we applied a quality threshold of 30 and ignored any base pairs
with quality less than the threshold. Table \ref{table:cpu} gives an overview
of the computational resources required for each data set. \SS{Additionally,
number of $k$-mers used by different methods for each data set is available 
in \ref{fig:entth}.}

\begin{figure*}[t]
\centering
\includegraphics[width=\textwidth]{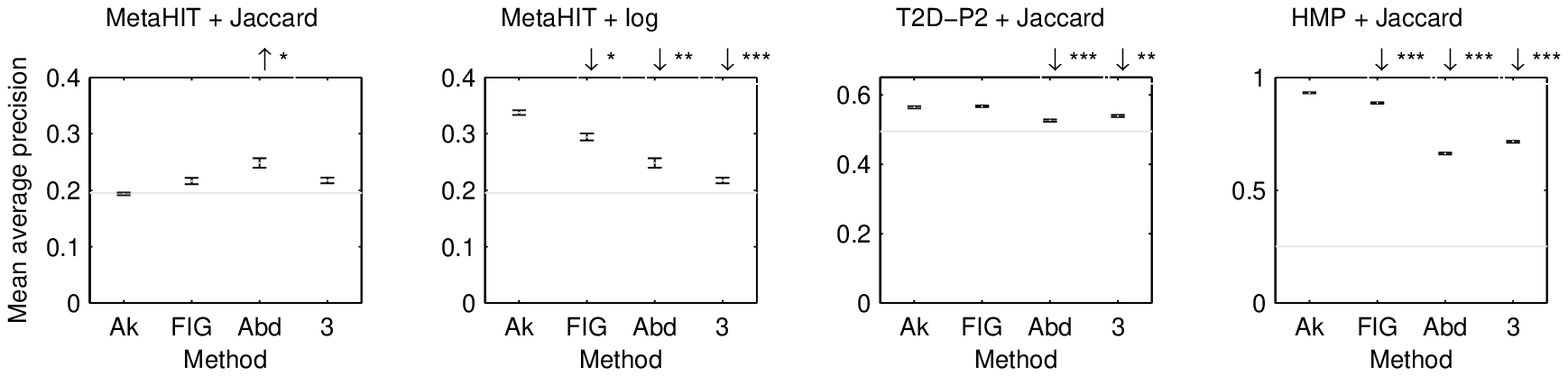}
\centering
\includegraphics[width=\textwidth]{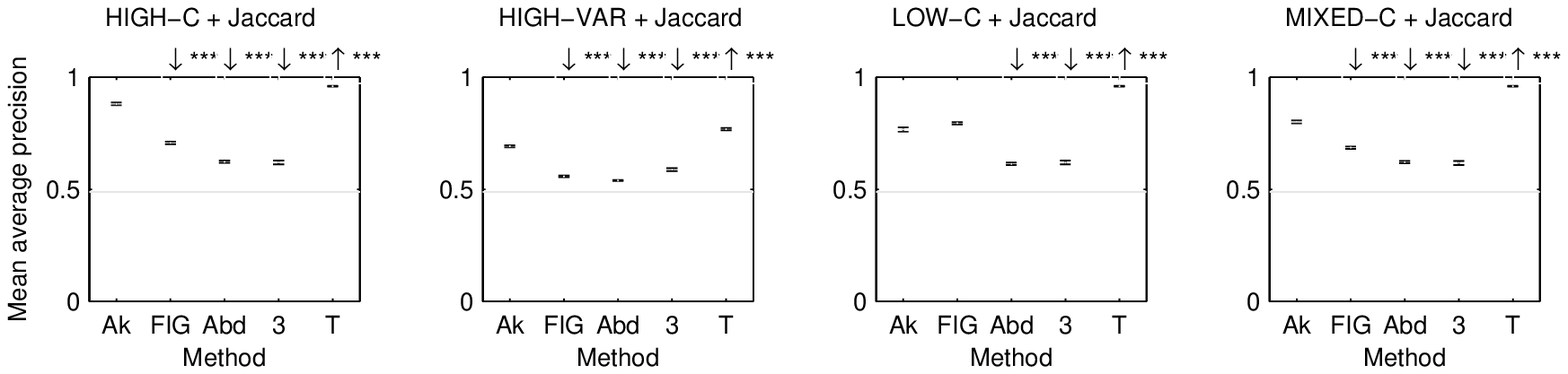}
\caption{Retrieval performance comparison of the proposed approach
  using all $k$-mers (``Ak'') against the following base measures:
  (1) ``FIG'': retrieval performance using known protein family,
  (2) ``Abd'': Hellinger distance between relative estimated
  abundance, (3) ``3'': $d^2_S$ distance between relative
  abundance of $3$-mers. \SS{``Ak'' uses the `optimized metric' over 101
  equally spaced threshold values between 0 and 1}. Each errorbar
  shows the MAP value along with the standard error. \SS{The grey
  horizontal line shows retrieval by chance: MAP computed over zero
  similarity metric}.  An
  arrow (if present) over a method indicates whether the performance
  of the corresponding method is significantly better ($\uparrow$) or
  worse ($\downarrow$) than ``Ak'': The stars denote significance
  level: 0 $<$ *** $<$ 0.001 $<$ ** $<$ 0.01 $<$ * $<$ 0.05.
  \SS{ For the
  synthetic datasets (in the bottom row) the relative abundance is known from
  experimental design. We present this result as ``T''. For MetaHIT
  we present the performance for both Jaccard and log metric, since the latter
  performs much better compared to the former.}}
\label{fig:compbase}
\end{figure*}

\begin{figure*}[t]
\centering
\includegraphics[width=0.75\textwidth]{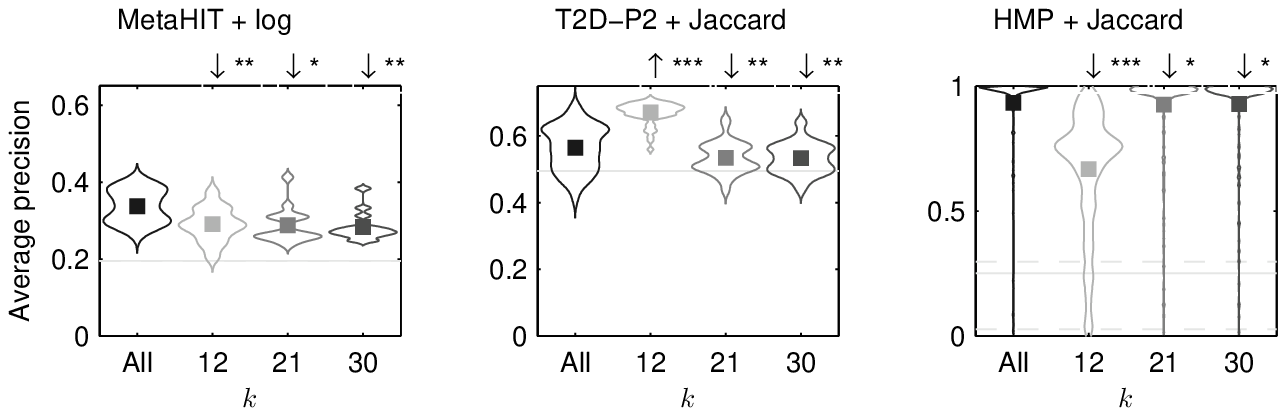}
\includegraphics[width=\textwidth]{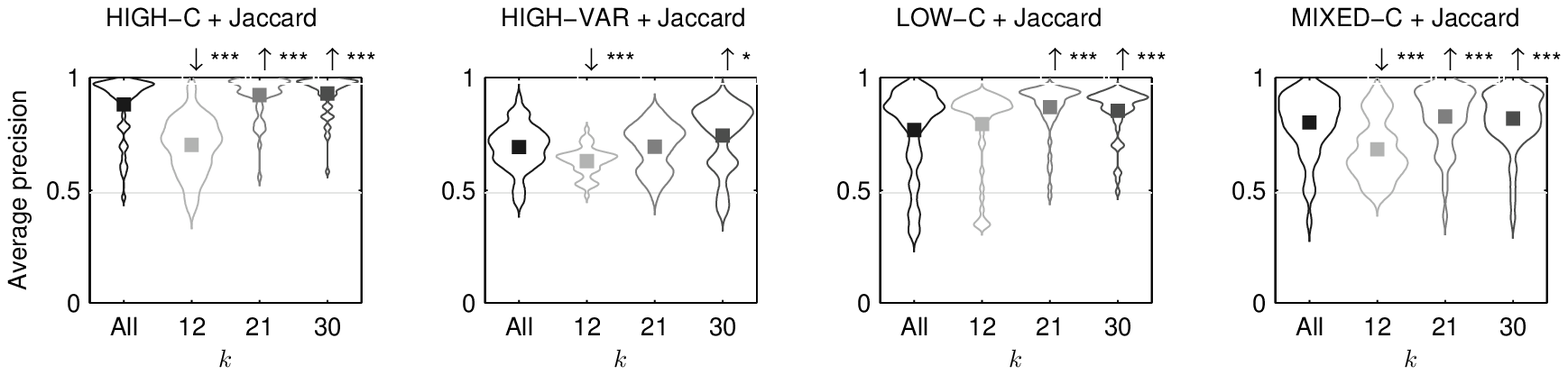}
\caption{Comparison of best performances for different $k$-mer
  lengths. The figures show the performance over queries by all
  positive samples as a violin plot. \SS{All methods use the `optimized metric' 
  chosen over 101
  equally spaced threshold values between 0 and 1}: the box denotes the MAP
  value. 
\SS{The horizontal lines show retrieval by chance: AveP computed over zero
dissimilarity metric. 
Straight line is the mean, and dotted lines are $5\%$, and $95\%$ quantiles
respectively, when number of relevant samples differ for different queries.}
  An arrow (if
  present) over a method implies whether the corresponding method
  performs significantly better ($\uparrow$) or worse ($\downarrow$)
  than `All' : The stars denote significance level: 0 $<$ *** $<$
  0.001 $<$ ** $<$ 0.01 $<$ * $<$ 0.05.  We observe that the
  considering all $k$-mers usually perform equally well with respect
  to considering a single $k$.}
\label{fig:basic}
\end{figure*}

\paragraph{Retrieval of samples with similar annotation:} \SS{
We applied the proposed approach and a number of alternatives to retrieval of
similar samples from the same data set and evaluated by how many of the
retrieved samples had the same annotation: class label, disease state or 
body site.  A comparison of the
obtained mean average precision values averaged over queries by all positive
samples is shown in Fig.~\ref{fig:compbase}. The results show the 
performance achieved by the `optimized metric'.  The alternatives we
considered were: i. retrieval performance based on the proposed distances but with
the frequencies counted on specific 21-mers from known protein families (FIGfams)
\cite{meyer_figfams:_2009}; ii. retrieval based on Hellinger distances between
relative species abundances estimated using \SS{MetaPhlAn}
\cite{segata_metagenomic_2012}; and iii. retrieval based on $d^2_S|M_0$ 
distances between relative frequencies of $3$-mers \cite{jiang_comparison_2012}.
}

\SS{
For the simulated data, the two classes differ only by the relative species
abundance; thus, retrieval based on ground truth abundance can be considered
to give an upper limit for the performance.
For HIGH-C and HIGH-VAR, the proposed method performs closer to 
the ground truth performance than any other methods, although 
the difference from ground truth performance is still statistically
significant. 
For LOW-C, the performance of all methods, except the protein family based
comparison, drop compared to HIGH-C,
while for MIXED-C the performance is again close to HIGH-C despite the presence
of low coverage samples. This is an encouraging observation showing the robustness of
the proposed approach to varying sequencing depths. 
}

\SS{For the real data sets the proposed approach yielded statistically
significantly higher mean average precision than any of the alternatives ($p <
0.05$) for all the datasets, except T2D-P2 where protein family based
comparison works equally well. 
Interestingly, the abundance-based retrieval performs relatively poorly
here, suggesting that the differences between the classes cannot be easily
captured by species composition alone, while the proposed $k$-mer features can
provide a better separation. Retrieval based on the known protein family performed
fairly well, but slightly worse than the proposed approach on MetaHIT. 
We observe that
for MetaHIT, Jaccard metric performs poorly; however, a change of metric to
log significantly improves the performance for all methods. Otherwise, all metrics usually
work equally well over different data sets.}

\paragraph{Effect of using specific or unspecific $k$-mer length:}
We next compared the proposed approach of using all $k$-mers to using a
specific $k$.  The retrieval performance using `optimized metric'
is shown in Fig.~\ref{fig:basic}.
\SS{
 The figures show the complete distribution
of average precision values over different queries whose mean is the mean
average precision of Fig.~\ref{fig:compbase}. The 
performance of the proposed method is usually better than with any individual $k$.
Thus, the proposed method appears to be a relatively safe choice that does
not suffer from catastrophically bad performance on any of the data sets.}

\begin{figure*}[t]
\centering
\includegraphics[width=0.75\textwidth]{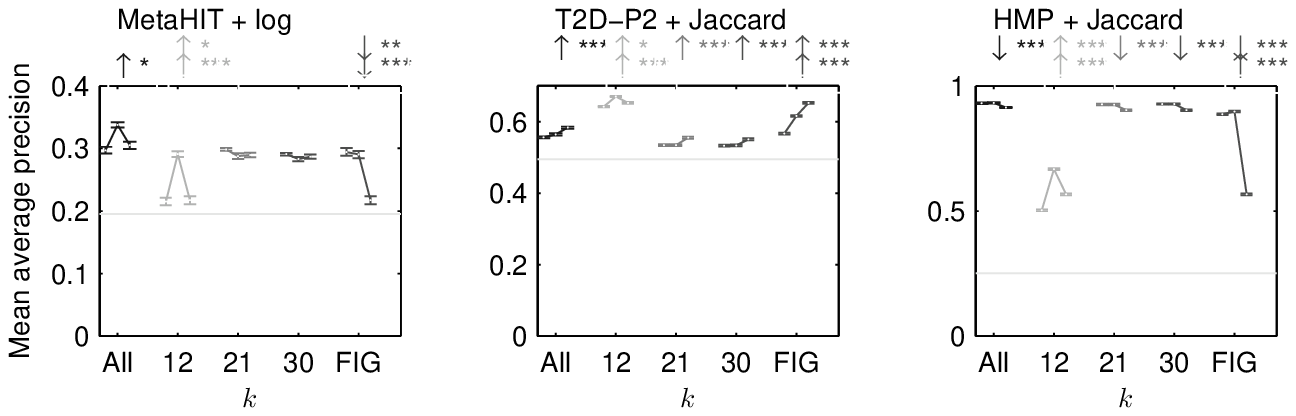}
\includegraphics[width=\textwidth]{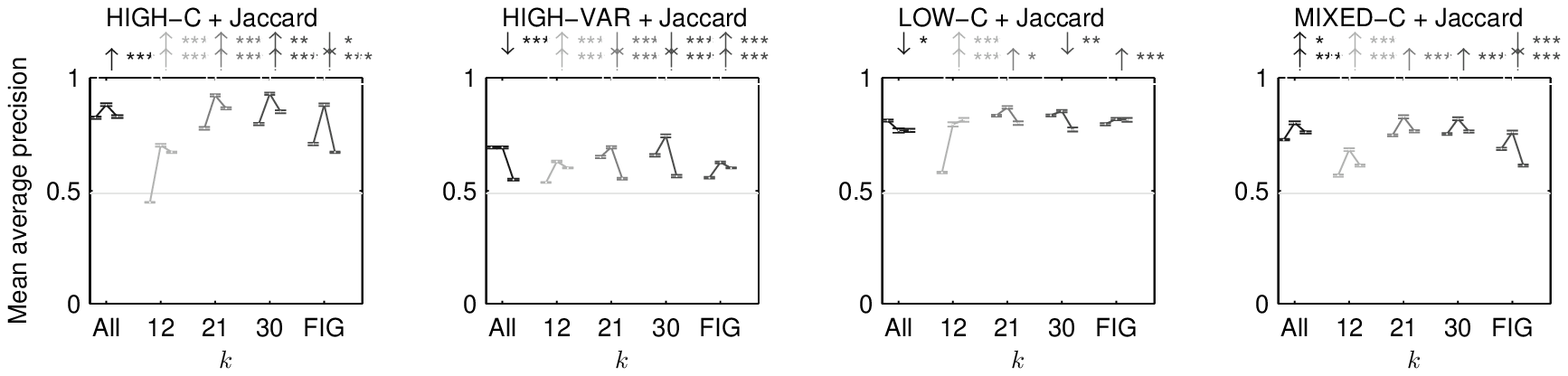}
\caption{\SS{Comparison of the best retrieval performance achieved with
`optimized metric' (middle), `average metric' (right) 
and without entropy filtering (left), for proposed approach `All', individual $k$s
as well as FIGfam based distance metric. The metrics are `optimized'/`averaged'
over 101 equally spaced threshold values between 0 and 1}.
 Each
errorbar line shows the MAP value along with the standard error.  
\SS{The grey horizontal line shows retrieval by chance: MAP computed
over zero dissimilarity metric.
An arrow (if present) over a method implies whether the performance of the
corresponding method (top: `average metric', bottom: `optimized metric') is
better ($\uparrow$) or worse ($\downarrow$) than}
 when entropy filtering is employed:
The stars denote significance level: 0 $<$ *** $<$ 0.001 $<$ ** $<$ 0.01 $<$ *
$<$ 0.05.  We observe that
filtering has a positive impact on the retrieval performance.}
\label{fig:compent}
\end{figure*}

\paragraph{Effect of the entropy filtering:}
Next, we evaluated the efficacy of filtering the informative $k$-mers against
retrieval performance without the filtering operation.  The results are
presented in Fig.~\ref{fig:compent}. We observed that 
entropy filtering usually improved retrieval performance
for all tested $k$-mer lengths \SS{when using the `optimized metric',
although the improvement might not always be statistically significant.
Although `average metric' often provides significant performance, it
might not always improve over performance without filtering.
Also, retrieval performance of FIGfam may or may not improve
with entropy filtering.}

\begin{figure*}[t]
\centering
\includegraphics[width=0.75\textwidth]{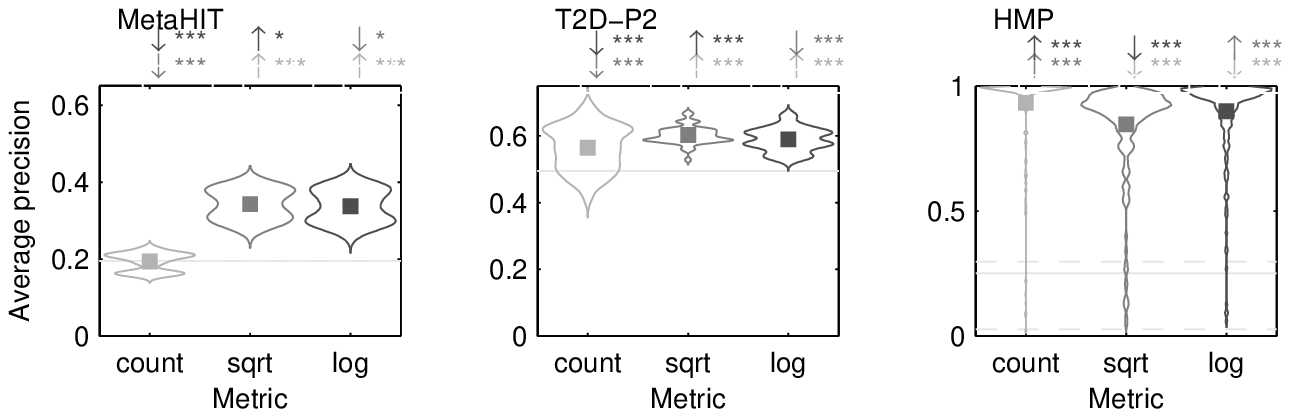}
\includegraphics[width=\textwidth]{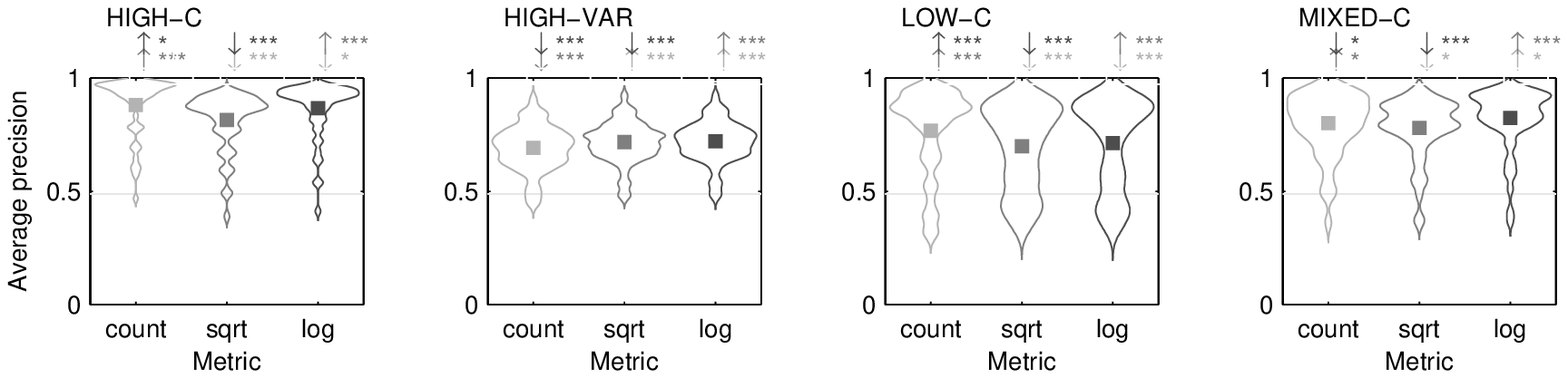}
\caption{Comparison of the best retrieval performance 
for different distance metrics using all $k$-mers. They show a
violin plot of the average performances over queries by all positive samples in
the data sets. \SS{The `optimized metrics' have been
selected over 101
  equally spaced threshold values between 0 and 1}:
the box denotes the MAP value. 
\SS{The horizontal lines show retrieval by chance: AveP computed over zero
dissimilarity metric. 
Straight line is the mean, and dotted lines are $5\%$, and $95\%$ quantiles
respectively, when number of relevant samples differ for different queries.}
  An arrow (if present) over a method implies whether the
corresponding method performs significantly better ($\uparrow$) or worse
($\downarrow$) than the other methods (denoted by their colors): The stars
denote significance level: 0 $<$ *** $<$ 0.001 $<$ ** $<$ 0.01 $<$ * $<$ 0.05.
We observe that different distance metrics usually demonstrate similar
performance.} \label{fig:compmethods}
\end{figure*}

\paragraph{Comparison across different metrics:}
Finally, we evaluated the retrieval performance over different dissimilarity
metrics. \SS{We presented the performance using `optimized metric'}
for different metrics in Fig.~\ref{fig:compmethods}.
We observed
that the simple presence/absence-based metric $D_{\mathrm{count}}$ performed at
least as well as abundance-sensitive log and sqrt metrics, \SS{except for the
MetaHIT data for which the other metrics performed the better}. 

%
%

\section{Conclusion}

In the wake of collecting multiple samples from similar environments
information retrieval for metagenomic samples is expected to become a handy
tool in metagenomics research.  In this paper, we have addressed the problem of
retrieving relevant metagenomic samples given a query sample from the same
collection.  The novelty of the proposed approach is that it is unsupervised,
and does not rely on the availability of reference databases.  We have
suggested employing $k$-mer frequencies as feature representation; however,
rather than exploring $k$-mers of a fixed $k$, we have scanned through all
possible $k$-mers of all possible $k$'s using distributed string mining, and
have proposed appropriate filtering technique to discard uninformative
$k$-mers. We have evaluated our method on both real and simulated data, and
observed that the approach can effectively retrieve relevant metagenomic
samples, outperforming both the FIGfams method based on known highly
informative protein families as well as retrieval based on species composition
of the samples.

\section*{Acknowledgement}
\SS{
The authors would like to thank Ahmed Sobih for his help with the
MetaPhlAn experiments on MetaHIT and T2D-P2.
Part of the calculations presented above were performed using computer
resources within the Aalto University School of Science ``Science-IT''
project.}
\paragraph{Funding:} 
This work was supported by the Academy of Finland (project numbers
140057, 250345, 251170 and 259440).


\end{document}